\documentclass[journal]{IEEEtran}

\newtheorem{DD}{Definition}

\ifCLASSINFOpdf
\else
\fi

\ifCLASSOPTIONcompsoc
\usepackage[nocompress]{cite}
\else
\usepackage{cite}
\fi

\usepackage{amsmath}
\usepackage{amssymb}
\usepackage{graphicx}
\usepackage{epstopdf}
\usepackage{mathtools}
\usepackage{xcolor}

\begin{document}

\title{Delay Doppler Transform}


\author{
        Xiang-Gen Xia, \IEEEmembership{Fellow}, \IEEEmembership{IEEE} 

\thanks{
  X.-G. Xia is with the Department of Electrical and Computer Engineering, University of Delaware, Newark, DE 19716, USA (e-mail: xxia@ece.udel.edu).
This work was supported in part by the National Science Foundation (NSF)  of USA under Grant CCF 2246917. }

}

\date{}

\maketitle


\begin{abstract}
  This letter is to introduce delay Doppler transform (DDT)
  for a time domain signal.
  It is motivated by the recent studies in wireless communications over
  delay Doppler channels that have both time and Doppler spreads, such as,
  satellite communication channels. We present some simple properties
  of DDT as well. The DDT study may provide insights of delay Doppler channels. 
\end{abstract}



\section{Introduction}\label{sec1}

The success of Starlink has re-generated world wide interest on satellite
communications. A special characteristic of satellite communications
is that its channel is not only time spread but also Doppler spread
for wideband transmissions. To deal with such channels, a recent popular
topic is orthogonal time frequency space (OTFS) modulation \cite{otfs1}
that has been shown identical to vector OFDM (VOFDM)  \cite{xia1, xia2}
in \cite{otfs2, otfs3, xia3, osdm}, at least, from the transmission side. 

For a delay Doppler channel, at time delay $\tau$, let
\begin{equation}\label{1.0}
h(\tau,t)=g(\tau) e^{-j\Omega(\tau)t}
\end{equation}
be its channel response with Doppler shift $\Omega(\tau)$. Let $s(t)$ be a transmitted signal. Then, the received signal $y(t)$ at time $t$  is
  \begin{eqnarray}
    y(t)& = & \int h(\tau,t)s(t-\tau)d\tau+w(t)  \nonumber\\
       & =& \int g(\tau) s(t-\tau) e^{-j\Omega(\tau)t} d\tau + w(t),\label{1.1}
  \end{eqnarray}
  where $w(t)$ is the additive noise.

  When the Doppler function $\Omega(\tau)$ in (\ref{1.1}) is
  a constant $\Omega$
  that does not depend on $\tau$, it means that all the channel responses
  at all the delays  have the same Doppler shift $\Omega$. In this case,
  this Doppler shift can be compensated at either transmitter or receiver
  and the compensated channel then becomes a time spread only channel.

\section{Definition}\label{sec2}

Motivated from the above delay Doppler channel model, we define delay
Doppler transform (DDT) below.

\begin{DD}\label{Def1}
Let $g(t)$ be a window function  and $s(t)$ be a signal.
The DDT of $s(t)$ is defined as
\begin{equation}\label{2.1}
  DDT_s(t, \Omega)=\int s(\tau) g(t-\tau) e^{-j\Omega(\tau)t} d\tau,
  \end{equation}
where $\Omega(\tau)$ is a  function of $\tau$.
\end{DD}

When function $\Omega(\tau)$ in (\ref{2.1}) is linear in terms of $\tau$,
i.e, $\Omega(\tau)=\Omega \tau$ for a constant $\Omega$,
the above definition becomes
\begin{equation}\label{2.2}
  DDT_s(t, \Omega)=\int s(\tau) g(t-\tau) e^{-j\Omega\tau t} d\tau. 
  \end{equation}
In this case, we call $\Omega$ as the Doppler shift rate (or frequency rate) of the
transform (or the channel).
The DDT in (\ref{2.2}) measures signal $s(t)$
by window function $g(-t)$ across its all time shifts and
Doppler shifts with a non-zero Doppler shift rate $\Omega$.
It is different from the short time Fourier transform (STFT) of $s(t)$
with window function $g(t)$, which is
\begin{equation}\label{2.20}
STFT_s(t, \Omega)=\int s(\tau) g(\tau-t) e^{-j\Omega\tau} d\tau.
\end{equation}
STFT is to measure $s(t)$ by a given window
function $g(t)$ across its all  time and frequency shifts.

When function $\Omega(\tau)$ in (\ref{2.1}) is constant, the above definition becomes
\begin{equation}\label{2.3}
  DDT_s(t, \Omega)=  e^{-j\Omega  t} \int s(\tau) g(t-\tau) d\tau, 
  \end{equation}
where $\Omega$ is  constant and does not depend on the time delay $\tau$.
We call it  a trivial Doppler spread. 
In this case, it is clear that after the compensation of the
common Doppler shift at transmitter, the DDT becomes the convolution, i.e.,
the channel is time spread only.

It is known that OFDM (or Fourier transform) converts a time spread only channel to multiple non-time spread subchannels. To improve the transmission signal
spectrum, a pulse (or window) with better spectrum than the
rectangular pulse is added, which is the generalized frequency division
multiplexing (GFDM) \cite{GFDM}. Another purpose to use a window
in GFDM is to limit the OFDM block size, i.e., to have a smaller block size
than the conventional OFDM to adapt to a time varying channel. 
GFDM is different from VOFDM (or OTFS),
unless the vector size in VOFDM is $1$ and then in this case, VOFDM returns
to OFDM. VOFDM converts a time spread only channel to
multiple vector subchannels where there is no time spread (or intersymbol interference (ISI)) across vector subchannels, while there is ISI inside
each vector subchannel. OFDM corresponds to discrete Fourier transform (DFT)
filterbank \cite{ppv}, VOFDM corresponds to vector DFT filterbank \cite{xia2},
and GFDM corresponds to  discrete Gabor transform (DGT) \cite{xia4} (or STFT with a given window function).

From (\ref{2.2}), the DDT, that corresponds to a non-trivial
 but simply a linear Doppler spread in terms of time delay, 
 is different from all the existing
 joint time-frequency transforms/distributions
 in the literature. This means that the existing joint time-frequency
 transforms may not be helpful to deal with non-trivial Doppler spread
 and time-spread channels. It also implies that it is not possible
 to well compensate  non-trivial Doppler spread at either transmitter
 or receiver.
So, neither GFDM nor VOFDM (OTFS) can compensate 
 a non-trivial Doppler spread well. 

 From (\ref{1.0}), one can see that the delay Doppler channel response
 function is not the most general two dimensional delay Doppler channel
 response function $h(\tau, t)$. 
It is even more difficult to compensate the Doppler spread in
a more general two dimensional delay Doppler channel.

For the inverse DDT, when the window function $g(t)$ is known,
$s(t)$ can be obtained from the deconvolution of $DDT_s(t, \Omega)$
at $\Omega=0$.

 \section{Properties}\label{sec3}

 We now present some simple properties of the DDT in (\ref{2.2}).
 We first consider the DDT of a time  shifted signal  $s(t-t_0)$
 with time shift $t_0$. 
Then, from (\ref{2.2}) it is not hard to see
 \begin{equation}\label{3.1}
   DDT_{s(t-t_0)}(t, \Omega)
   = DDT_{s(t)e^{-j\Omega t_0t}}(t-t_0, \Omega) e^{-j\Omega t_0 t}.
 \end{equation}
 We know that
 the Fourier transform or STFT of a time shifted signal
 is that of the orignal signal modulated in frequency. However, 
 from (\ref{3.1}), one can see that it is different from the Fourier transform
 or STFT 
 in the sense that
 the DDT of a time shifted
 $s(t)$ is the time shifted and additionally
 modulated DDT of the modulated $s(t)$.

 For the delay and Doppler channel (\ref{1.1}), the received signal
 can be represented by the DDT of the transmitted signal $s(t)$
 with the channel response amplitude function $g(t)$  as
 the window function below:
 \begin{equation}\label{3.2}
   y(t)=DDT_s(t,-\Omega) e^{-j\Omega t^2} +w(t).
 \end{equation}
 From (\ref{3.2}), one can see that the signal part of the received
 signal is the linear chirp modulated  DDT of the transmitted signal
 evaluated at the negative Doppler shift rate, i.e., $-\Omega$.
 In other words, the dechirped received signal $y(t)e^{j\Omega t^2}$
 is a DDT of the transmitted signal. This implies that the study of DDT
 is important for the communication over the delay Doppler channel (\ref{1.1}).

 We next consider a transmitted signal in a communication system: 
 \begin{equation}\label{3.3}
   s(t)=\sum_n s_n p(t-nT),
 \end{equation}
 where $s_n$ are the information symbols to transmit, $p(t)$ is the pulse,
 and $T$ is the symbol duration. Then, using the property (\ref{3.1}),
 its DDT is 
 \begin{equation}\label{3.4}
   DDT_s(t, \Omega) =  \sum_n s_n DDT_{p_n}(t-nT, \Omega)e^{-j n T t \Omega},
 \end{equation}
 where $p_n$ is the modulated $p(t)$: 
 \begin{equation}\label{3.40}
 p_n=  p_n(t)=p(t)e^{-jnT \Omega t}.
   \end{equation}
 From (\ref{3.2}) and (\ref{3.4}), 
 at the receiver we have the following new channel:
 \begin{equation}\label{3.5}
y'(t)=\sum_n s_n DDT_{p_n}(t-nT, -\Omega)e^{j n T t \Omega}+
   w'(t),
 \end{equation}
 where $y'(t)=y(t) e^{j\Omega t^2}$ and  $w'(t)=w(t) e^{j\Omega t^2}$. 
 Since the above dechirping is a unitary operation, it does not change
 the received signal or the noise property. The above DDT based
 receive signal model (\ref{3.5})
 might provide insights in designing pulses $p(t)$ in better dealing with
delay Doppler channels in communications systems. 

 \section{Simulations}\label{sec4}
 We now see some plots of the DDT in (\ref{2.2})  and STFT in (\ref{2.20})
 for some simple signals.
 The window function we use is a Gassian function
 $g(t)=e^{-t^2}$. Three signals are tested.
The first is a linear chirp $s_1(t)=e^{jt^2}$, the second is a quadratic chirp
$s_2(t)=e^{jt^3/10}$, and the third is their sum, i.e.,
$s_3(t)=s_1(t)+s_2(t)$. All of them are supported on $[-10,10]$. 
The magnitudes of DDT and STFT of these three signals
 in the region $(t, \Omega)\in [-5,5]\times [-5, 5]$ are shown in Figs. 1-6.

 We know that the STFT roughly tells the joint time frequency
 distribution property for a signal, although its resolution may
 not be as high as those of non-linear time frequency
 distributions \cite{qian}, such as Wigner-Ville
 distribution.  From these figures, we find that the DDT
 of a signal is much different from a joint time frequency distribution,
 which may help to understand a delay Doppler channel more.

\begin{figure}
\centering
\includegraphics[scale=0.6]{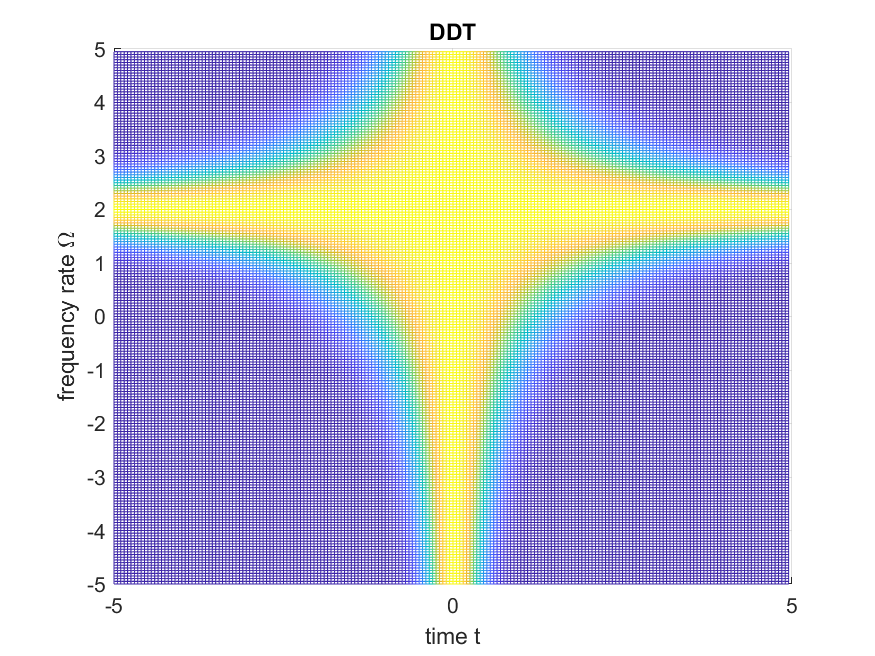}\\
(a)\\
\includegraphics[scale=0.6]{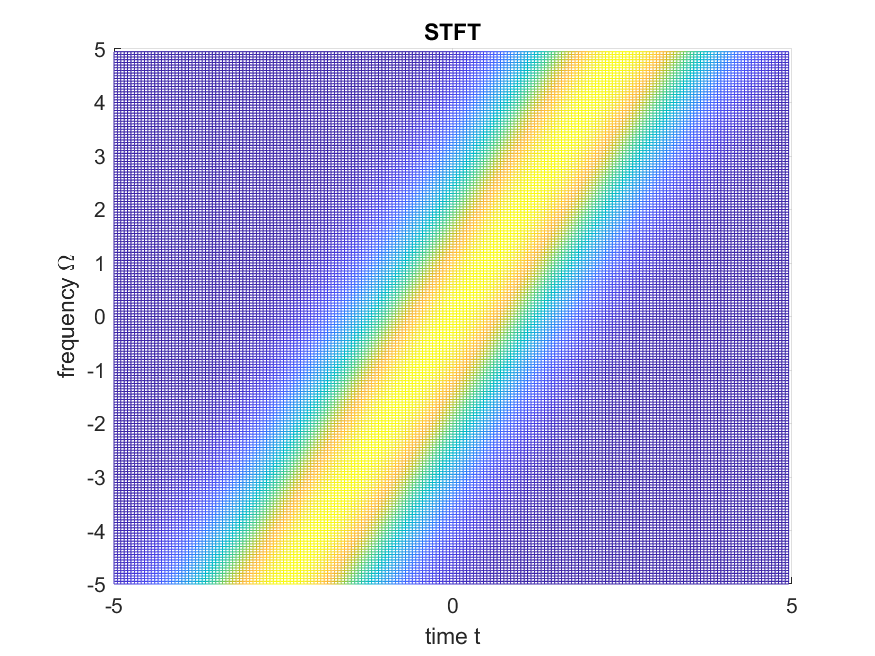}\\
(b)

\caption{The DDT and STFT of signal $s_1(t)=e^{jt^2}$: (a) DDT; (b) STFT.}
\label{fig1}
\end{figure}

\begin{figure}
\centering
\includegraphics[scale=0.6]{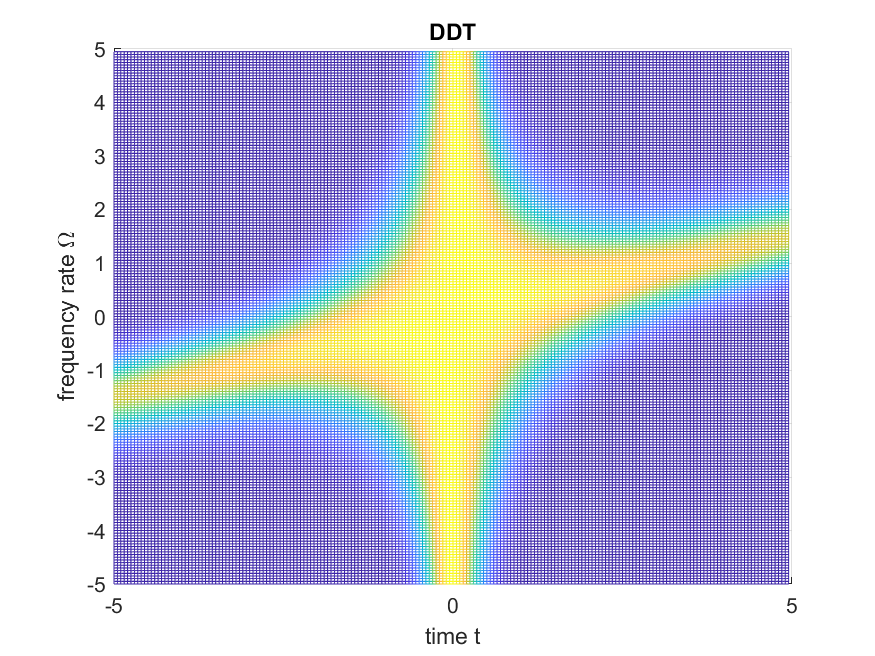}\\
(a)\\
\includegraphics[scale=0.6]{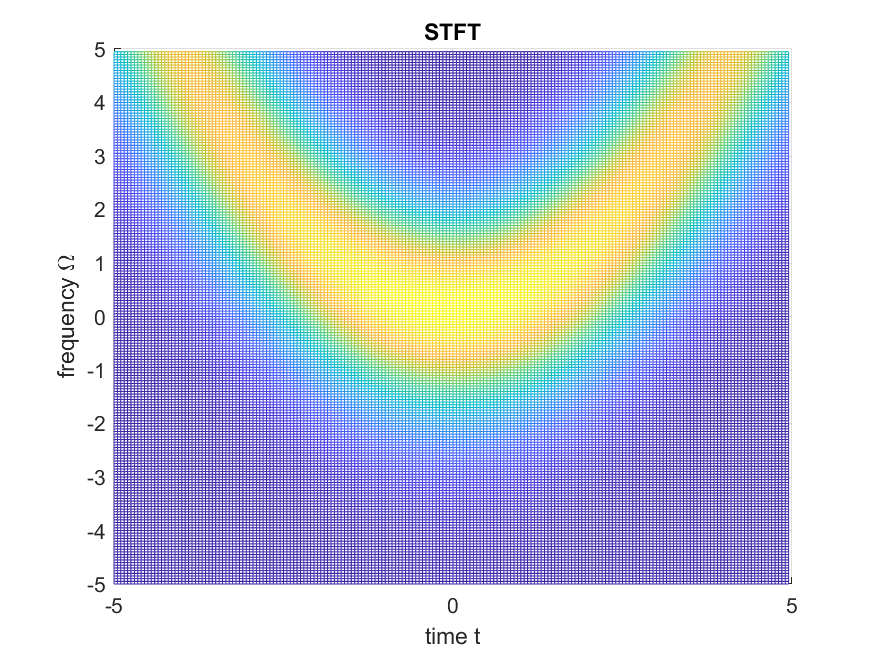}\\
(b)

\caption{The DDT and STFT of signal $s_2(t)=e^{jt^3/10}$: (a) DDT; (b) STFT.}
\label{fig2}
\end{figure}

\begin{figure}
\centering
\includegraphics[scale=0.6]{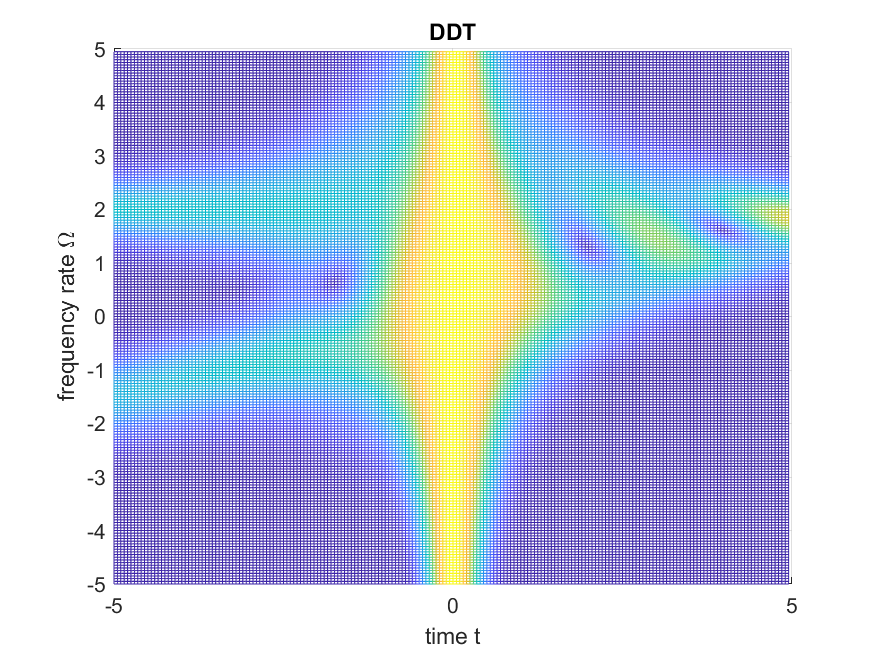}\\
(a)\\
\includegraphics[scale=0.6]{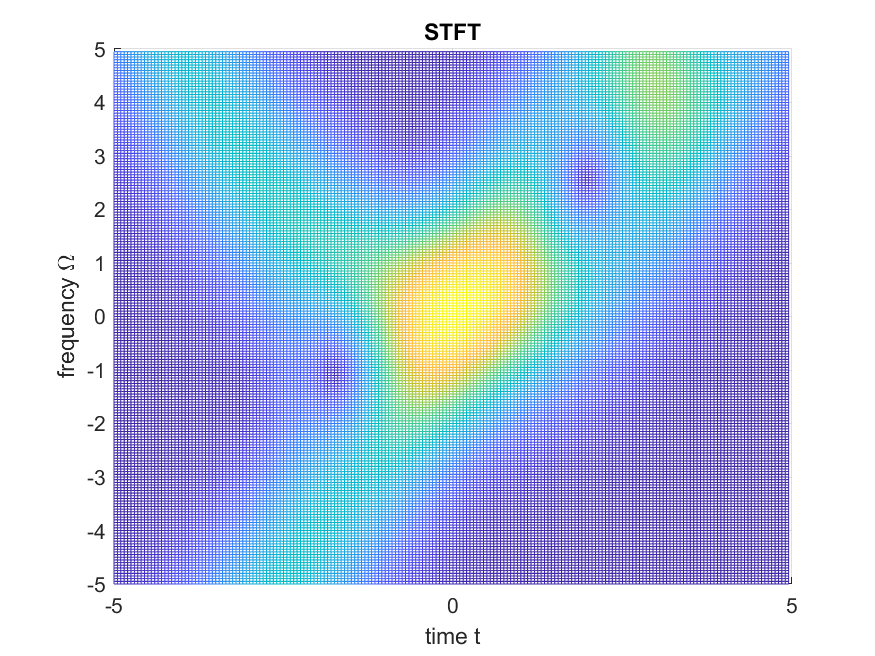}\\
(b)

\caption{The DDT and STFT of signal $s_3(t)=s_1(t)+s_2(t)=e^{jt^2}+e^{jt^3/10}$: (a) DDT; (b) STFT.}
\label{fig3}
\end{figure}

\section{Conclusion}
In this letter, we introduced delay Doppler transform (DDT) for a signal.
It was movitaed from the recent interest in wireless communications over
delay Doppler channels, such as satellite channnels. We
also provided some simple properties about DDT. One can see that
DDT is different from all
the existing joint time frequency analysis techniques and may provide
more insights for delay Doppler channels.

\end{document}